\documentclass[manuscript]{jkas}
 \usepackage{ulem}
\def\year{yyyy} % publication year
\def\month{mm} % publication month
\def\volume{00} % journal volume
\def\issue{0} % journal issue
\def\beginpage{00} % first page of article
\def\endpage{00} % last page of article
\def\received{mm dd, yyyy} % date paper was received by JKAS
\def\accepted{mm yy, yyyy} % date of acceptance
\date{Received \received; accepted \accepted}

\title{
Toward Next Generation Solar Coronagraph: Diagnostic Coronagraph Experiment
}

\author[1,2]{K.-S. Cho \thanks{The corresponding author of this document.}}
\author[1]{H. Yang}
\author[1]{J.-O. Lee}
\author[1]{S.-C. Bong}
\author[1]{J. Kim}
\author[1]{S. Choi}
\author[1]{J.-Y. Park}
\author[3]{K.-H. Cho}
\author[1]{J.-H. Baek}
\author[1]{Y.-H. Kim}
\author[1]{Y.-D. Park}

\affil[1]{Korea Astronomy and Space Science Institute, Daejeon, 305-348, Republic of Korea; \email{kscho@kasi.re.kr}}
\affil[2]{University of Science and Technology, Daejeon, 305-330, Korea}
\affil[3]{Astronomy Program, Department of Physics and Astronomy, Seoul National University, Seoul 151-747, Republic of Korea}

\def\corrauthor{
K.-S. Cho
}

\def\runningauthor{
K.-S. Cho
}

\def\runningtitle{
Cho et al.
}

\def\keywords{
Coronagraph, Solar Corona, Electron temperature, Electron velocity, Solar wind, International Space Station
}

\def\abstracttext{
Korea Astronomy and Space Science Institute (KASI) has been developing a next-generation coronagraph (NGC) in cooperation with NASA to measure the coronal electron density, temperature, and speed simultaneously using four different filters around 400 nm. KASI organized an expedition team to demonstrate the coronagraph measurement scheme and the instrumental technology through the 2017 total solar eclipse (TSE) across the USA. The observation site was in Jackson Hole, Wyoming, USA. We built an eclipse observation system, so-called Diagnostic Coronal Experiment (DICE), which is composed of two identical telescopes to improve a signal to noise ratio. The observation was conducted with 4 wavelengths and 3 linear polarization directions according to the planned schedule in a limited total eclipse time of about 140 seconds. Polarization information of corona from the data was successfully obtained but we were not able to obtain global information of coronal electron temperature and speed in the corona due to a low signal-to-noise ratio of the optical system and a strong emission from the prominence located in the western limb. In this study, we report the development of DICE and observation results from the eclipse expedition. TSE observation and analysis by using our own developed instrument gave an important lesson that a coronagraph should be carefully designed to achieve the scientific purpose of this study. And it was a very useful experience in the way for the success of follow-up NASA-KASI joint missions called the Balloon-borne Investigation of the Temperature and Speed of Electrons in the Corona (BITSE) and COronal Diagnostic EXperiment (CODEX).
}

\begin{document}
\jkashead %% set title, authors, abstract, etc.

%%%%%%%%%%%%%%%%%%%%%%%%%%%%%%%%%%%%%%%%%%%%%%%%%%%%%%%%%%%%%%%%%%%%%
%%% BEGIN MAIN TEXT HERE %%%%%%%%%%%%%%%%%%%%%%%%%%%%%%%%%%%%%%%%%%%%
%%%%%%%%%%%%%%%%%%%%%%%%%%%%%%%%%%%%%%%%%%%%%%%%%%%%%%%%%%%%%%%%%%%%%
\section{Introduction\label{sec:intro}}

A solar corona is the outermost region of the Sun's atmosphere that consists of a low density magnetized plasma with a temperature of a few million degrees. It extends millions of kilometers into space and is the source of continuous outflows of solar wind. Observations of the corona have been conducted by using white light coronagraphs that adopt an occulting disk to block the Sun's bright surface, allowing the faint solar corona to be seen. So far, ground-based white light coronagraphs such as the Mark-IV K-coronameter (MK-4) and the COronal Solar Magnetism Observatory (COSMO) K-coronagraph (K-Cor) on the Mauna Loa Solar Observatory (MLSO), and space-based white light coronagraphs such as Large Angle Spectroscopic Coronagraph \citep[LASCO;][]{brueckner1995} onboard the Solar and Heliospheric Observatory (SOHO) and Sun-Earth Connection Coronal and Heliospheric Investigation \citep[SECCHI;][]{howard2008} onboard the Solar TErrestrial Relations Observatory (STEREO) have contributed to understanding the structure and evolution of the solar corona and the origin of CMEs, as well as their interactions with the Earth.

Expanding the capability of currently operating coronagraph, the next generation coronagraph (NGC) is expected to enable studies on the plasma structure of solar corona, dynamics of accelerating solar wind, and CME kinematics by determining the electron density, temperature, and velocity simultaneously in the distance range $\sim$ 3 - 8 $R_\odot$ simultaneously \citep{cho2017}. Main purpose of the NGC is to find signatures of hot plasma, and to give answer to the question on the velocities and temperatures of density structures that are observed ubiquitously within streamers and coronal holes. The observation wavelength is approximately 400 nm, where strong Fraunhofer absorption lines from the photosphere experience thermal broadening and Doppler shift through scattering by coronal electrons. The strong absorption lines are flattened by the thermal motions of free coronal electrons, and their profiles depend on coronal temperatures. Therefore, the electron temperature can be determined by using the amount of flatness of the absorption lines, which can be represented by the ratio of intensities in two temperature-sensitive wavelengths. Moreover, the radial motion of solar wind affects the red-shift of the entire coronal spectrum. The ratio of two speed-sensitive wavelength intensities also facilitates the estimation of the speed of the solar wind \citep{cram1976, reginald2001, cho2016, reginald2017}. Based on this observation method, the Korea Astronomy and Space Science Institute (KASI) and NASA's Goddard Space Flight Center (GSFC) have been developing a NGC whose technology was demonstrated through the balloon experiment called the Balloon-borne Investigation of the Temperature and Speed of Electron in the Corona (BITSE) launched in September 2019 in Fort Sumner, New Mexico. Final purpose of this joint development is to install the NGC on the International Space Station (ISS) in 2023 through the COronal Diagnostic EXperiment (CODEX).

A total solar eclipse (TSE) is a useful experiment to understand the principle of the NGC observation in the early phase of the collaboration between KASI and NASA for the coronagraph development. For this, we developed an eclipse observation system, so-called Diagnostic Coronal Experiment (DICE) that is an optical camera system including the compact lens with a diameter of 50 mm, filter wheels, polarizer wheels, and CCD systems. By using the DICE, we conducted test observation through the North-American TSE expedition in 2017 to verify fundamental technology on optics, filter system, and operating system. This paper is about the development of the DICE and the observation by the instrument. It is organized as follows.

In Section 2, we describe the development of optics, filter system, detector, and operating system. In Section 3 and 4, we present instrument tests and TSE observation. The results and discussion are given in Section 5 and Section 6.

\section{DICE development\label{sec:Obs}}
For the 2017 TSE observation, two identical optical systems were installed on the single mount as shown in Figure 1. Each system operates with a bandpass filter unit, a polarizer unit, and a camera on a small computer. One of the DICE units is shown in the upper panel of Figure 1. It is composed of an optical lens group, a bandpass filter unit, polarizer unit, and a CCD camera. In front of the lens group, the baffle is equipped to block any stray light from outside field angle, and a pin-hole cover is placed to calibrate the intensity brightness of the solar corona and protect lenses and the CCD camera while the DICE is not operating. Two units
are mounted on a computer controlled motorized tracking system.

%with a sequence described in Table 3 for the observation.

\begin{figure}
\centering
\includegraphics[width=80mm]{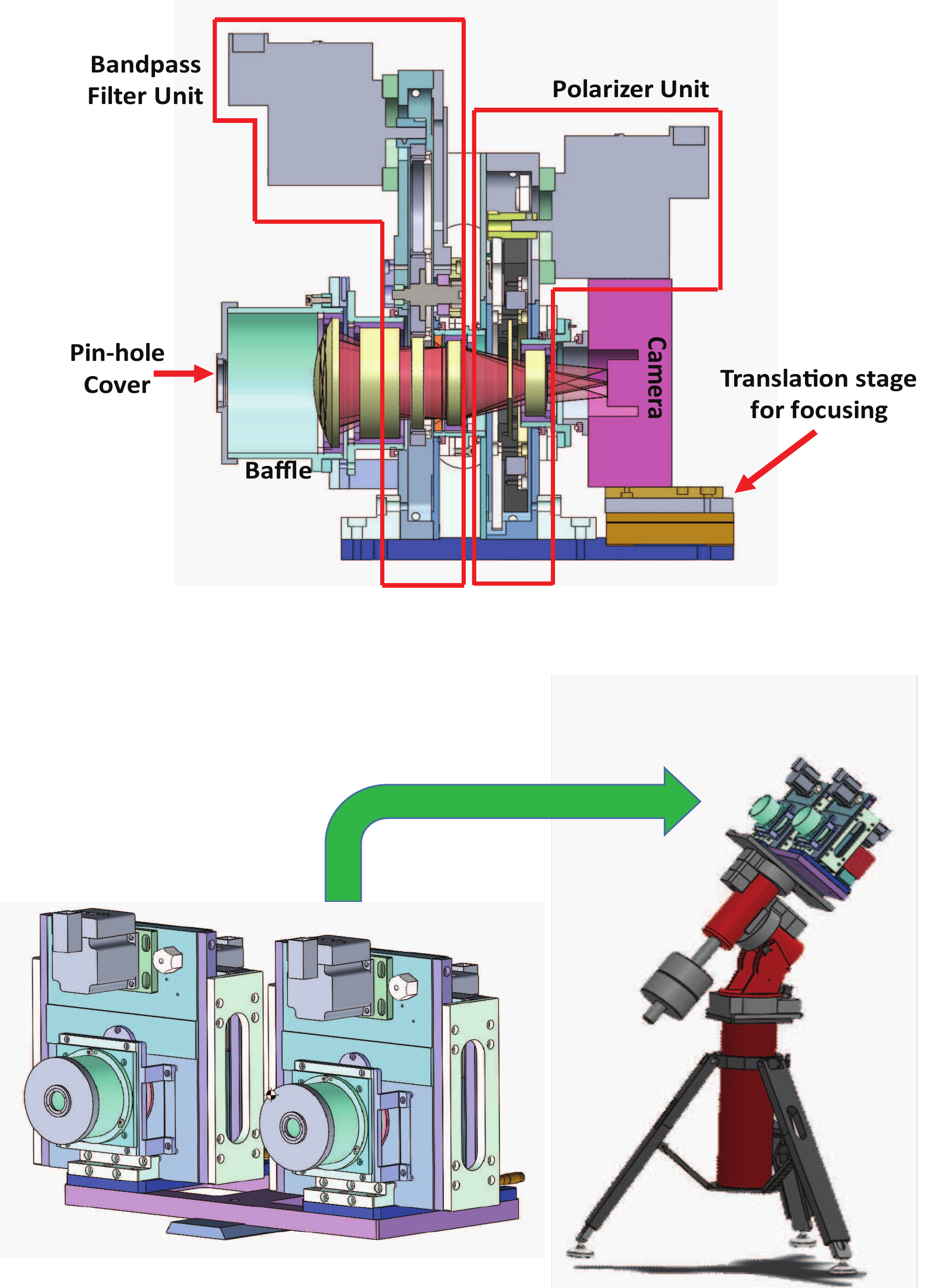}
\caption{Overall layout of DICE (upper panel) and two DICEs that are mounted on a tracking system (lower panel).}
\label{overall_layout}
\end{figure}

\subsection{Optics Design}
In order to provide successful observation of the TSE, the optical system has several requirements. The major requirements are system resolution and accuracy of the measured intensity. To achieve the system resolution, the RMS spot diameter of the optical system should be less than two pixels of the CCD whose pixel size is 7.4 $\mu$m, thus the RMS spot diameter should be less than 14.8 $\mu$m. For the measured intensity accuracy, the intensity variation due to stray light or polarization angle offset should be less than $\pm$ 1$\%$. Table 1 summarizes the requirements of the overall optical system.

\begin{table*}[t]
\centering
\caption{Optical design requirements for DICE optical system}
\label{tab:signif_gap_clos}
\begin{tabular}{cc}
\hline
Parameter & Requirement \\
\hline\hline
Half Field Of View (HFOV) & 1 $R_\odot$ $\sim$ 14$R_\odot$ (0.227$^\circ$ $\sim$ 3.73$^\circ$)  \\
\hline
Entrance pupil diameter & 50 mm  \\
\hline
System wavelength & 385 nm $\sim$ 410 nm  \\
\hline
Filter wavelengths & 393.9, 402.5 nm (Temperature) \\
                   & 399.0, 424.9 nm (Speed)  \\
\hline
 & QSI 640 4.2mp Cooled CCD Camera  \\
CCD camera & Sensor size: 15.5 $\times$ 15.5 mm  \\
 & 1 pixel = 7.4 $\mu$m $\times$ 7.4 $\mu$m  \\
\hline
\end{tabular}
\end{table*}

The optical design works are carried out based on the requirements. Beside those specifications, the design works should consider two plane windows for bandpass and polarization filters. Figure 2 shows the final optical design which consists of 4 lenses and 2 filters. Each surface of the lens is a spherical surface except the back surface of the lens 2. Using the aspherical surface on the lens 2 provides an advantage for simplifying the overall system. In the past time for the optical designing works, the aspheric surface is rarely used due to fabricating and testing difficulties. However, recently those works are not challenging in these days. The stop is located at the center of the lens group to reduce odd aberrations. As shown in Figure 2, the distance between the last surface of the lens 4 and CCD is about 30 mm, which allows a space for CCD camera moving to find the best focusing position.

\begin{figure}
\centering
\includegraphics[width=77mm]{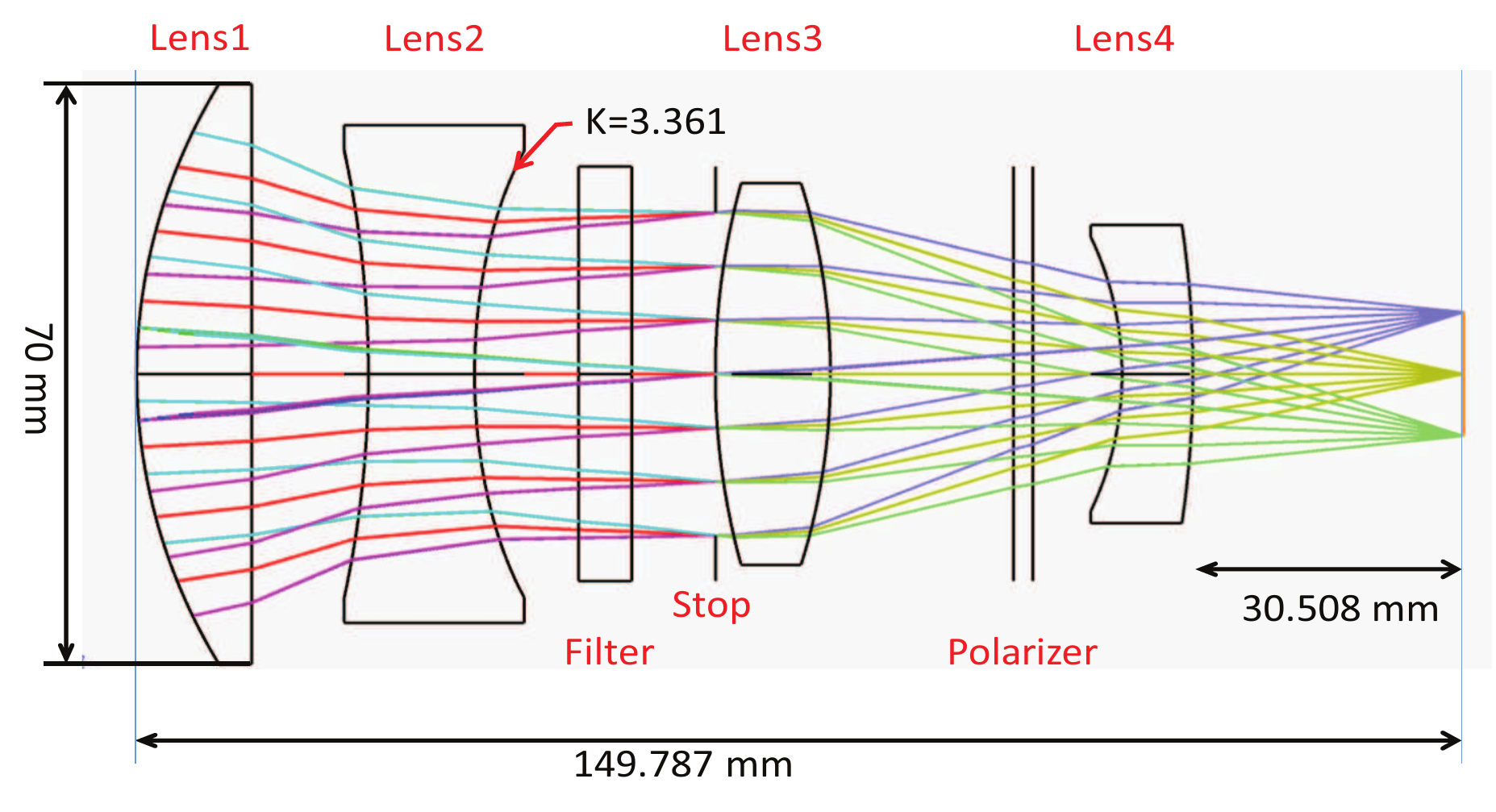}
\caption{Optical layout of the DICE. It consists of 4 lenses, filter, and polarizer. All lens surfaces are spherical except the back surface of the lens 2. }
\label{opt_layout}
\end{figure}

Figure 3 is a spot diagram to show the expected performance of the system. The RMS diameter is about 5.6 $\mu$m which is much less than 14.8 $\mu$m of two-pixel sizes. The different value between nominal design and the requirement would be a margin for other parameters such as fabricating, manufacturing, and aligning errors which are usually difficult to achieve the nominal requirements.

\begin{figure}
\includegraphics[width=108mm]{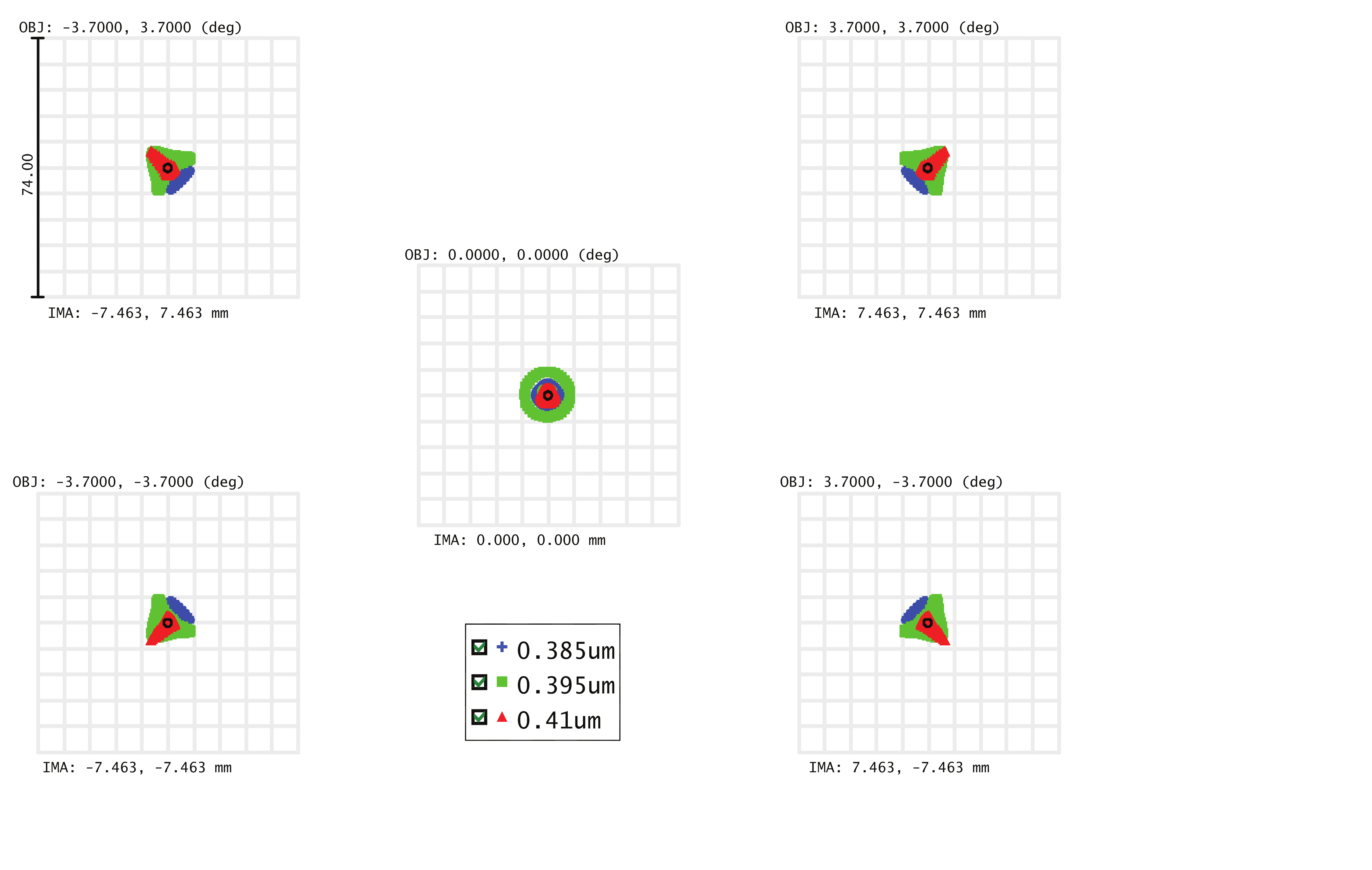}
\caption{Spot diagram of the optical design for different wavelengths (0.385, 0.395, and 0.41 $\mu$m).} RMS spot diameter at the outermost field is 5.6 $\mu$m which is less than the requirements.
\label{spot_dig}
\end{figure}

\begin{figure}
\centering
\includegraphics[width=80mm]{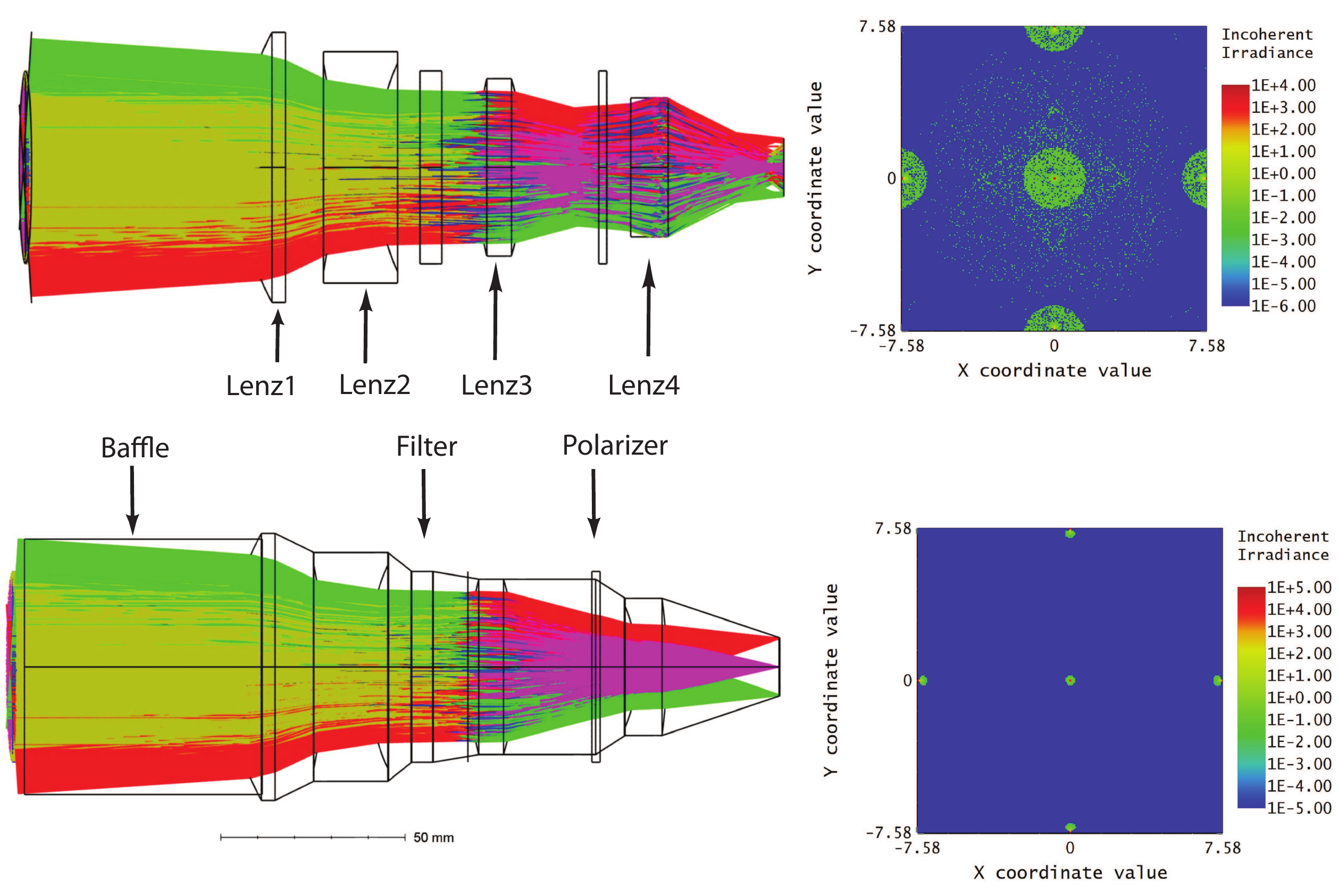}
\caption{Ghost analysis before (upper) and after (lower) AR-coating on the polarization filter. The ray-tracing shows the polarizer causes double bounces and ghost images. The total power on the CCD is 3.4755 Watts. The AR coated polarizer does not make the ghost images. The total power on the CCD is 3.6174 Watts.}
\label{ghost}
\end{figure}

The tolerance processing is done by Zemax, ray-tracing software package for the optical design works. In that process, all parameters are perturbed and analyzed for the performance. The tolerance for each optical element is simulated with random perturbation to see if the performance still meets the requirements. Most of the tolerance values are reasonable if the lens maker fabricates in the specified precision level. One of the difficulties to fabricate the lenses is the aspheric surface in the back surface of the lens 2. The tolerance of the conic constant is $\pm$0.0125, and the performance still meets the requirements with that error. The lens 2 was fabricated by a local lens manufacturer in Korea who uses Magnetorheological Finishing (MRF) for the finishing process and Aspheric Stitching Interferometer (ASI) for testing surface error.

To reduce the first-order scatter light, we take a ghost focus analysis that is related to the lens group and baffles. We perform the analysis to understand how much unwanted photon energy from Fresnel refraction and reflection of the lens interface reaches on the CCD surface. The DICE is optimized at the 405nm of the central wavelength, and all lenses are anti-reflection (AR) coated in wavelength from 385nm to 410nm for 10$^\circ$ of incident angle such that the reflection against the lens surface is less than 0.7\% of the incident light. We take the analysis in Zemax to find any critical optical elements causing any unexpected beam on the CCD. As a result, we find that the filters consisting of the flat surfaces causes double bouncing, and ghost focus is imaged on the CCD. It is also confirmed that the polarization filters, if not AR coated, produce some amount of ghost effects on the CCD (Figure 4). After applying AR coating on the polarizer, we can achieve less than 1\% of the photon energy loss due to ghost image, which indicates that the ghost effect for the bandpass filter is minor. In fact, the bandpass filters could not be AR coated since AR-coating on the surface reduces bandpass performance.

\subsection{Filter and Polarizer system}

The bandpass filters are purchased from Andover custom bandpass filters for each center wavelength with 10 nm of bandwidth, and blocking in other wavelengths up to far infrared. We use the Thorlabs' WP50L-UB which is ultra broadband wire grid polarizer with AR coating. Originally it comes with no coating on both surfaces, and its transmission and extinction ratio are $\sim$ 70$\%$ and $\sim$ 5000 respectively. We apply AR coating on the polarizer to reduce ghost effects as shown in Figure 4. The total powers are 3.4755 and 3.6174 out of 5 watts of total incoming power before and after the AR coating on the polarizer, respectively. The difference is about 0.14 watt or about 4\% of noise reduction. Figure 5 shows the filter and polarizer units.

\begin{figure}
\centering
\includegraphics[width=80mm]{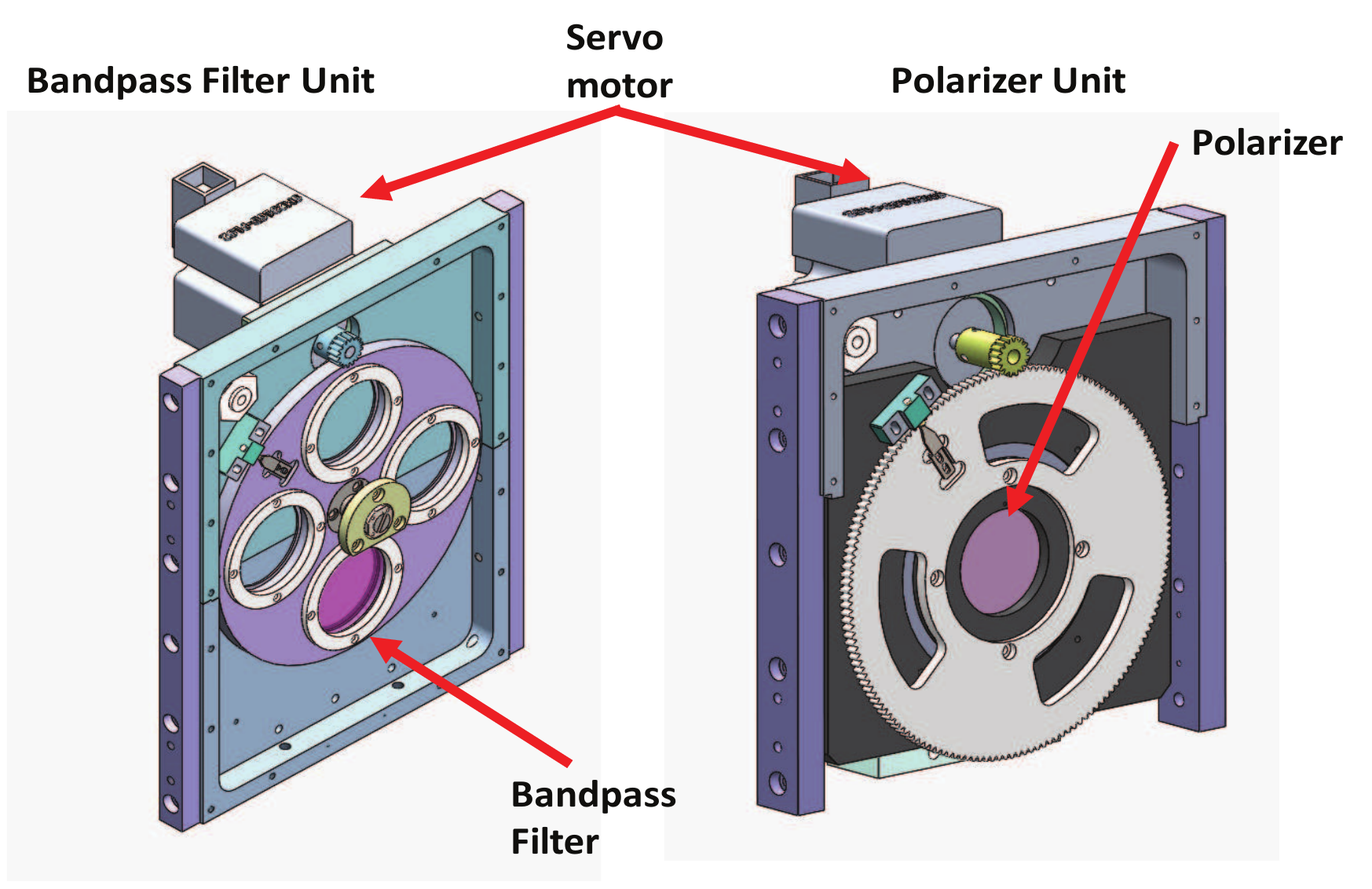}
\caption{Filter and polarizer units}
\label{filter_polarizer_units}
\end{figure}

\subsection{Baffle design}
Another factor that reduces the system performance is the light coming from outside of the field of view of the system. In the simulation, those unwanted light produces a little noise on the CCD. Limiting the bundle of the light with 64.5mm of a baffle in front of the first lens can remove those noise. The baffles are also the mount of the lens and filter. We used Lambertian scatter model, which is an ideal diffuser surface, for the scattering model of the baffle surface. We compared the CCD view with and without front baffles, and found that the difference between with and without the baffle is about 0.02\% of noise introduced.

\section{Instrument Test\label{sec:Test}}
In order to guarantee the performance of the system, the proper tests for each requirement are mandatory.

\subsection{Resolution and Stray light test}
The requirement for the resolution of the DICE is to make the RMS spot diameter smaller than two CCD pixels. During the DICE system assembly and test process, a CCD camera was not ready yet. Thus Thorlabs' DCU22M CCD camera with 4.65 $\mu$m of the pixel size was used for the test. For the test setup, we placed a white light point source at the long-distance and imaged it on the CCD through the DICE optics. As a result, we found that the image of the point source in the maximum intensity occupies two pixels or 9.3 $\mu$m, which is less than the requirement of 14.8 $\mu$m. Although the spot is too small to identify, the spot shape looks elliptical, which means the system has a little aberration. It is also found that the image also does not show any noticeable noise. The maximum peak value of the point source image is 255 with noise level 18.1$\pm$ 6.12.

\subsection{Polarizer Angle Test}
The other requirement is to make the polarization angle accurate so that the intensity variation is less
than 1\% in the same orientation in every rotation. During the TSE observation, the orientation of the polarizer rotates every 60$^\circ$, and the image in each angle is used to obtain polarization information such as polarization brightness and degree of linear polarization. For the test of the polarization angle accuracy, a linearly polarized 405nm laser source is used with the polarizer in the DICE. The intensity measurements are carried out with 10 revolutions of the polarizer. As shown in Figure 6, the standard deviation is 0.68\% which is less than 1\%.

\begin{figure}
\centering
\includegraphics[width=80mm]{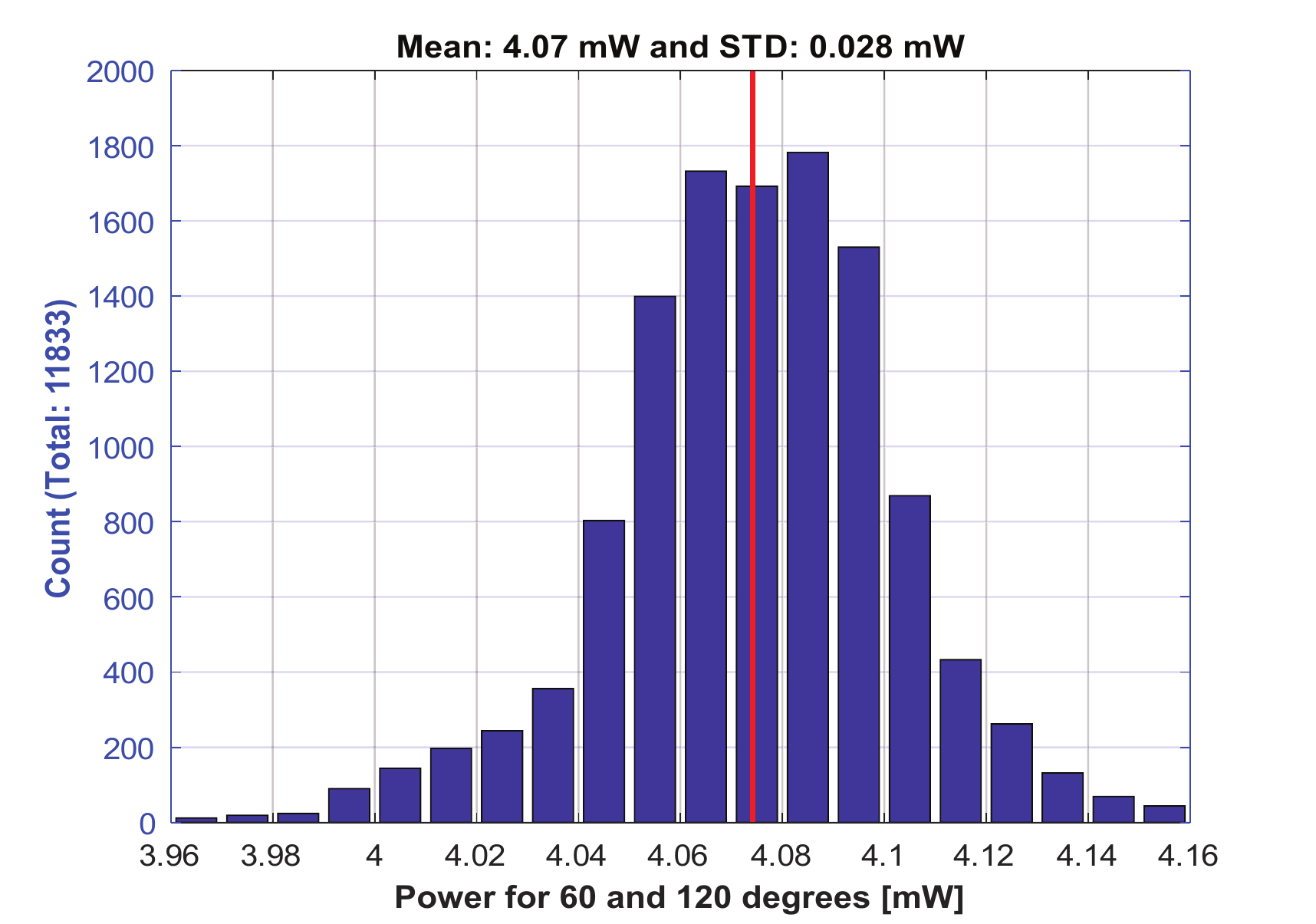}
\caption{Statistic result of the polarizer angle accuracy test. Standard deviation is 0.68\% which is less than requirement (1\%).}
\label{pol_rot_test}
\end{figure}

\subsection{Operation system and Data acquisition }

A computer used in the DICE control system is ASUS UN65U model, which has 2.5GHz of clock speed, 4GB of program memory for the operating system and control software, and 256GB of mass storage for observation data. We installed Ubuntu 16.04 LTS 32bit on this computer to build an operating environment for the DICE control software. The computer supports USB3 and Ethernet ports as an interface for communication with peripherals. The filter wheel connects two motors with an RS-232 interface in series using USB-RS232 converters, and the camera connects to a USB2.0 compliant interface. The two operation systems use a separate Ethernet port to connect to the wireless Ethernet switch and establish a network environment that allows them to connect the user's computer via a wireless connection.

We use the core Flight System (cFS) of NASA to develop DICE control software as shown in Figure 7.
The cFS is a reusable software framework for real-time systems especially in space missions \citep{mccomas2016}. The DICE control software is aimed at the successful observation of the TSE. The control software is configured in a layered architecture as shown in the upper panel of Figure 7. We developed four mission applications consisting of the camera control (CAM), two filter wheel control (WHL), science data processing and storing (DAP), and observation automation (OBS). With the mission applications, the DICE control software was built integrating three open-source applications of command ingestion (CI), telemetry output (TO), and scheduler (SCH) for periodic command generation. The Lower panel of Figure 7 shows the graphical user interface of the DICE remote control software. The remote control software sends and receives command and telemetry packets through UDP communication with the control software running on the DICE control computer. The remote control software was developed to control the filter wheel and camera using various parameters and to display observation progress and down-sampled images. It was utilized from the DICE assembly and alignment to the TSE observation.

\begin{figure}
\centering
\includegraphics[width=80mm]{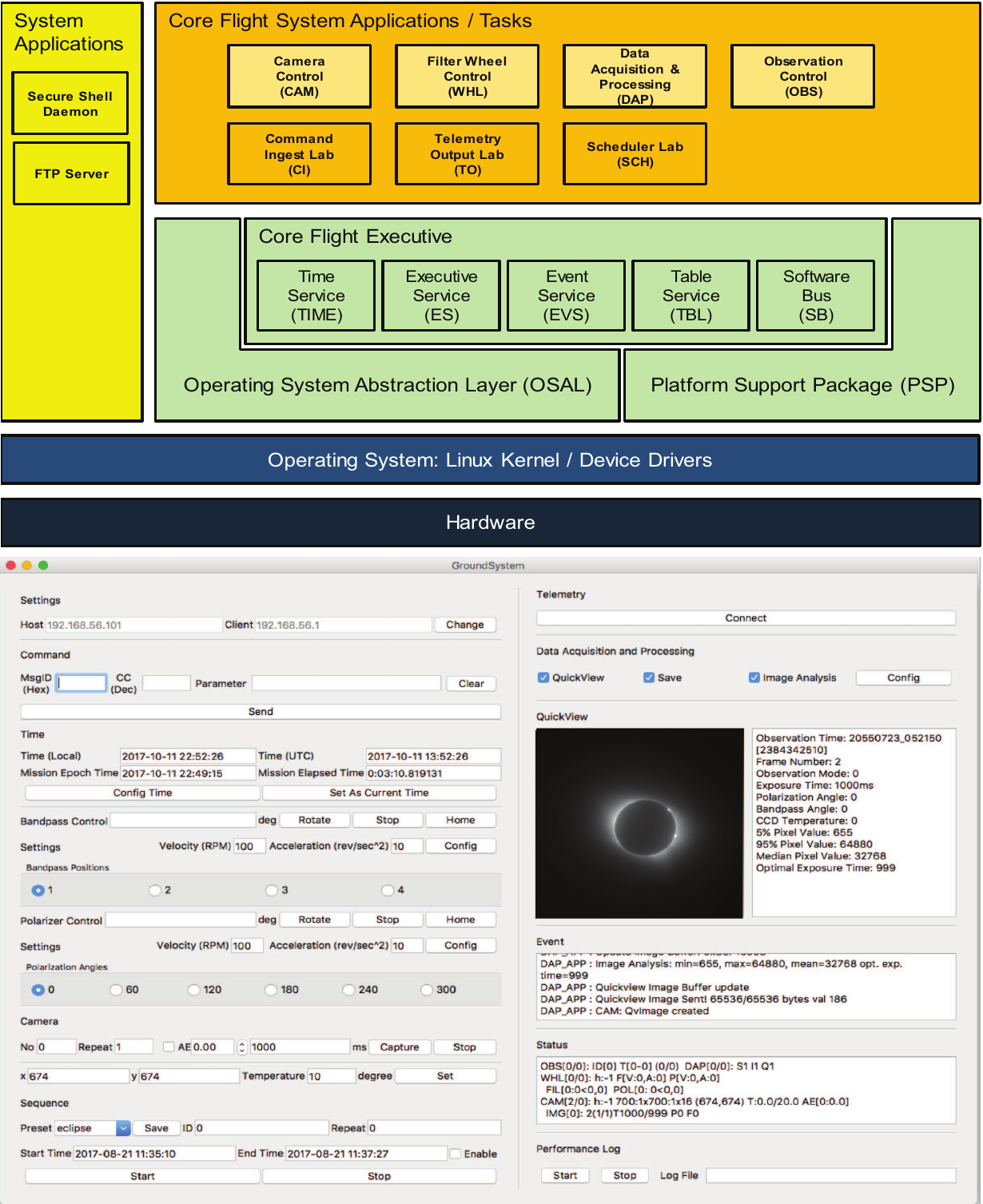}
\caption{DICE control system. Upper panel shows the layered architecture of DICE control software and lower panel present the graphic user interface of the DICE remote control software.}
\label{dice_control_system}
\end{figure}

\subsection{Overall System Tolerance and Image Stacking Method\label{sec:tol}}

Figure 8 shows the transmissions of the optical elements with the quantum efficiency of the CCD ($QE_{ccd}$) and the calculated coronal irradiance at 1.2R$_{\odot}$. $QE_{ccd}$ is measured using a spectrograph with integration sphere with 2\% of measuring tolerance. The bandpass filter transmission ($Tr_{fil}$), the polarizer transmission ($Tr_{pol}$), and the extinction ratio of the polarization $ext_{pol}$ is provided by manufacturer, Andover and Thorlabs, with a tolerance of 2.5\%, 1\%, and 1\% of transmission and extinction variation. We assumed the tolerance of the bandpass filter central wavelength as 2.5\,\AA~by considering the transmission tolerance of 2.5\% of the filter with 100\,\AA~bandwidth. The lens transmission ($Tr_{lens}$) is simulated using Zemax software. We assumed the tolerance of $Tr_{lens}$ as 1\%.

\begin{figure*}[t]
\centering
\includegraphics[width=\textwidth]{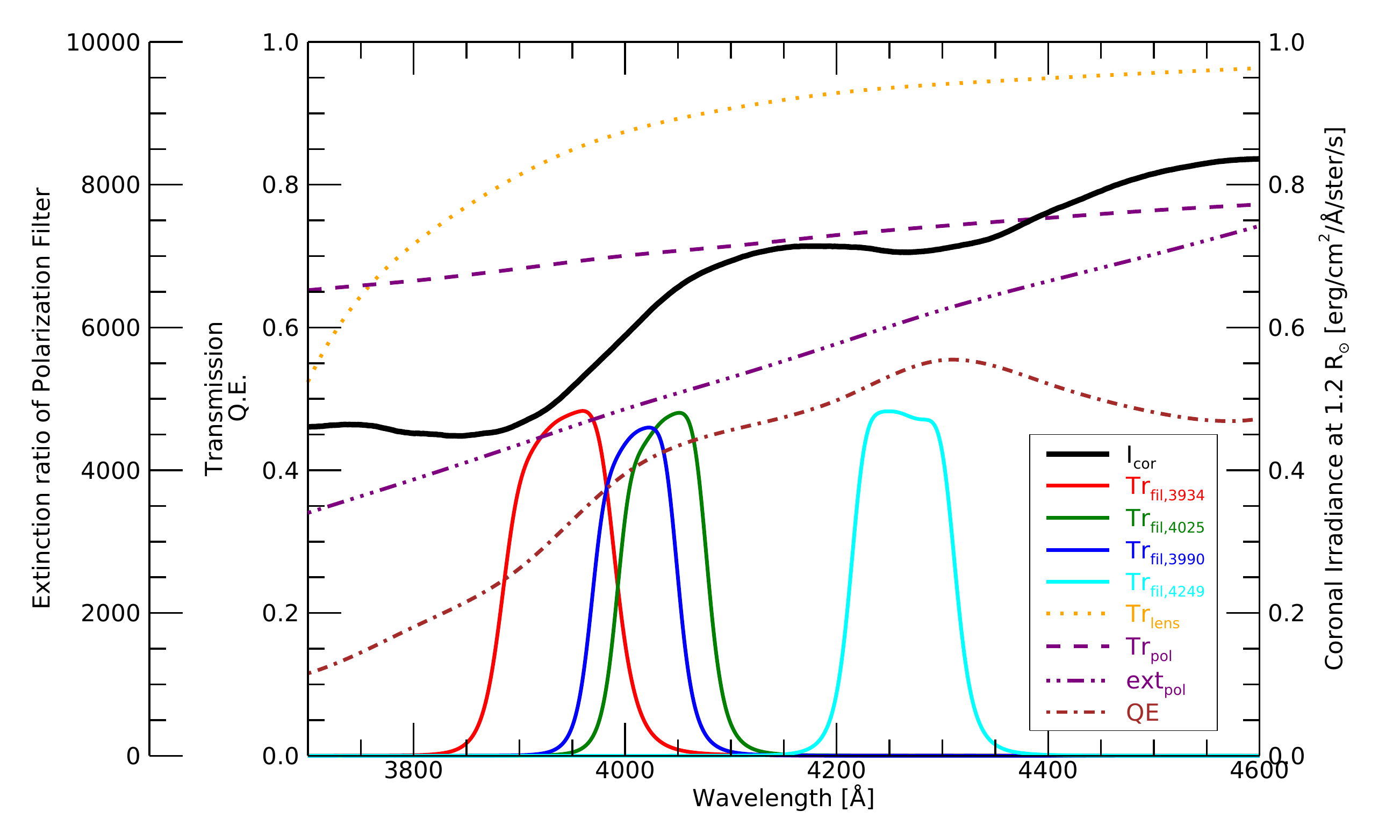}
\caption{Measured and simulated optical performance of the lens, bandpass filters, and CCD including the coronal irradiance (I$_{cor}$) at 1.2R in the condition of the electron temperautre of $10^6\,$K and the solar wind speed of $400\,$km/s.}
\label{tr_graph}
\end{figure*}

The signal recorded in a specific pixel of the CCD is given as,

\begin{eqnarray}\label{sigeqn}
 s_\theta&=&\int_\lambda [ \frac{Tr_{{fil}, \lambda}*Tr_{pol, \lambda}*Tr_{lens, \lambda}}{E_{ph, \lambda}}*Q_{ccd, \lambda}     \nonumber\\
 &*&A_{ap}(I_{t, \lambda}\cos^2\theta+I_{r, \lambda}\sin^2\theta+0.5*I_{unpol, \lambda})\nonumber\\
 &*&Omega_{pixel}/g*t_{exp} \nonumber\\
 &+&\frac{Tr_{fil, \lambda}*Tr_{pol, \lambda}/ext_{pol, \lambda}*Tr_{lens, \lambda}}{E_{ph, \lambda}}*Q_{ccd, \lambda}\nonumber\\
 &*&A_{ap}(I_{t, \lambda}\sin^2\theta+I_{r, \lambda}\cos^2\theta+0.5*I_{unpol, \lambda})\nonumber\\
 &*&\Omega_{pixel}/g*t_{exp}  ] \, d\lambda
 \end{eqnarray}

, where $E_{ph}$ is photon energy, $A_{ap}$ is the aperture size of the optics, $\Omega_{pixel}$ is the solid angular size per a pixel, $t_{exp}$ is the exposure time, $g$ is the gain of the camera, $\theta$ is the polarization filter angle with respect to the tangential direction, and $ext_{pol}$ is the extinction ratio of the polarization filter. $I_t$, $I_r$, and $I_{unpol}$ denote the tangentially and radially polarized brightness and the unpolarized brightness coming through the aperture, respectively. $I_t$ and $I_r$ contains K-coronal brightness, while $I_{unpol}$ contains the unpolarized light coming from sky and F-corona. $I_t$ and $I_r$ are described as the sum and subtraction of polarized brightness ($pB_{kcor}$) and the total brightness ($tB_{kcor}$) of the K-corona.
\begin{eqnarray}
I_t=0.5 (tB_{kcor}+pB_{kcor})\nonumber\\
I_r=0.5 (tB_{kcor}-pB_{kcor})
\end{eqnarray}

For the calculation, $tB_{kcor}$ and $pB_{kcor}$ are imported from the Thomson scattering calculations by assuming the isothermal corona and the coronal electron density model of \citet{baumbach1937}, in the condition of the electron temperature of 1\,MK and the solar wind speed of 400 km $s^{-1}$. The detailed calculation scheme is done following \citet{cho2016}. The second term of the equation \ref{sigeqn} represents the light polarized orthogonal to the polarizer which is not perfectly extinct, while the first term of the equation \ref{sigeqn} represents the polarized light passing through the polarizer.

\begin{figure}[t]
\centering
\includegraphics[width=0.5\textwidth]{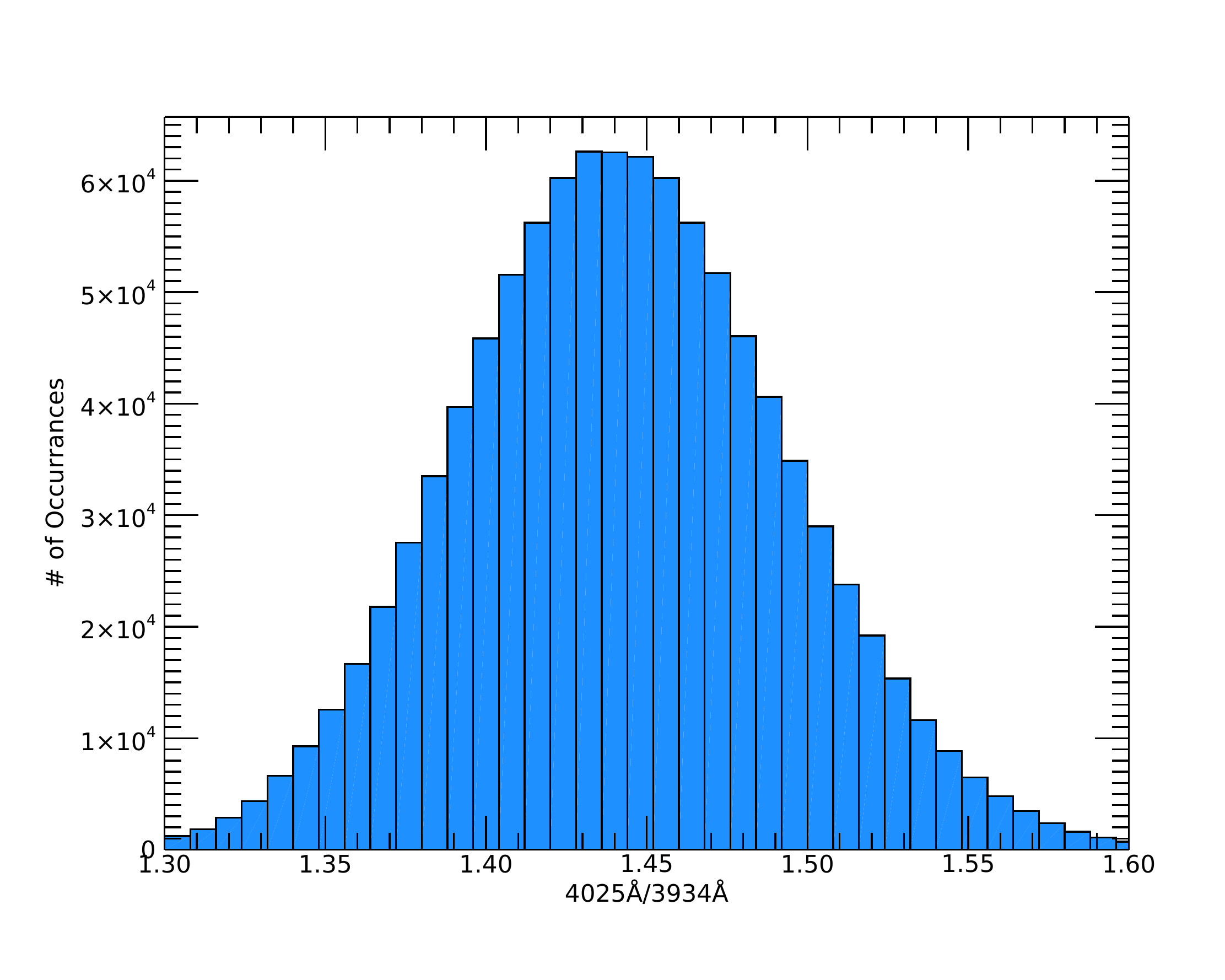}
\caption{Filter ratio scattered by the systematic tolerance. The ratio is calculated by applying the exposure times during the TSE observation (see Table 3). We also include the effect of the photon noise, dark, and readout to compare with the real observed ratio.}
\label{systol}
\end{figure}

Figure \ref{systol} presents the estimated filter ratio of the 4025\,\AA~and 3934\,\AA~ filters for temperature calculation. The ratio is calculated using the Monte-Carlo method, which is often used to model how the different errors propagate over a system. We generate the gaussian distributed random errors of optical and mechanical tolerance, photon noise, dark, and readout. We apply the exposure times given in Table 3, which were taken during the TSE observation. The random sampling is repeated for $10^6$ samples. The 3-$\sigma$ deviation from 1.29 to 1.59 centering at 1.44 of the value represents the systematic tolerance propagates on the observed result. For example, if we made a DICE with no error on the instrument and there was no noise during the observation, the filter ratio shall be 1.44. Otherwise, the ratio shall be shifted to obtain the deviating temperature.

\begin{figure}[t]
\centering
\includegraphics[width=0.5\textwidth]{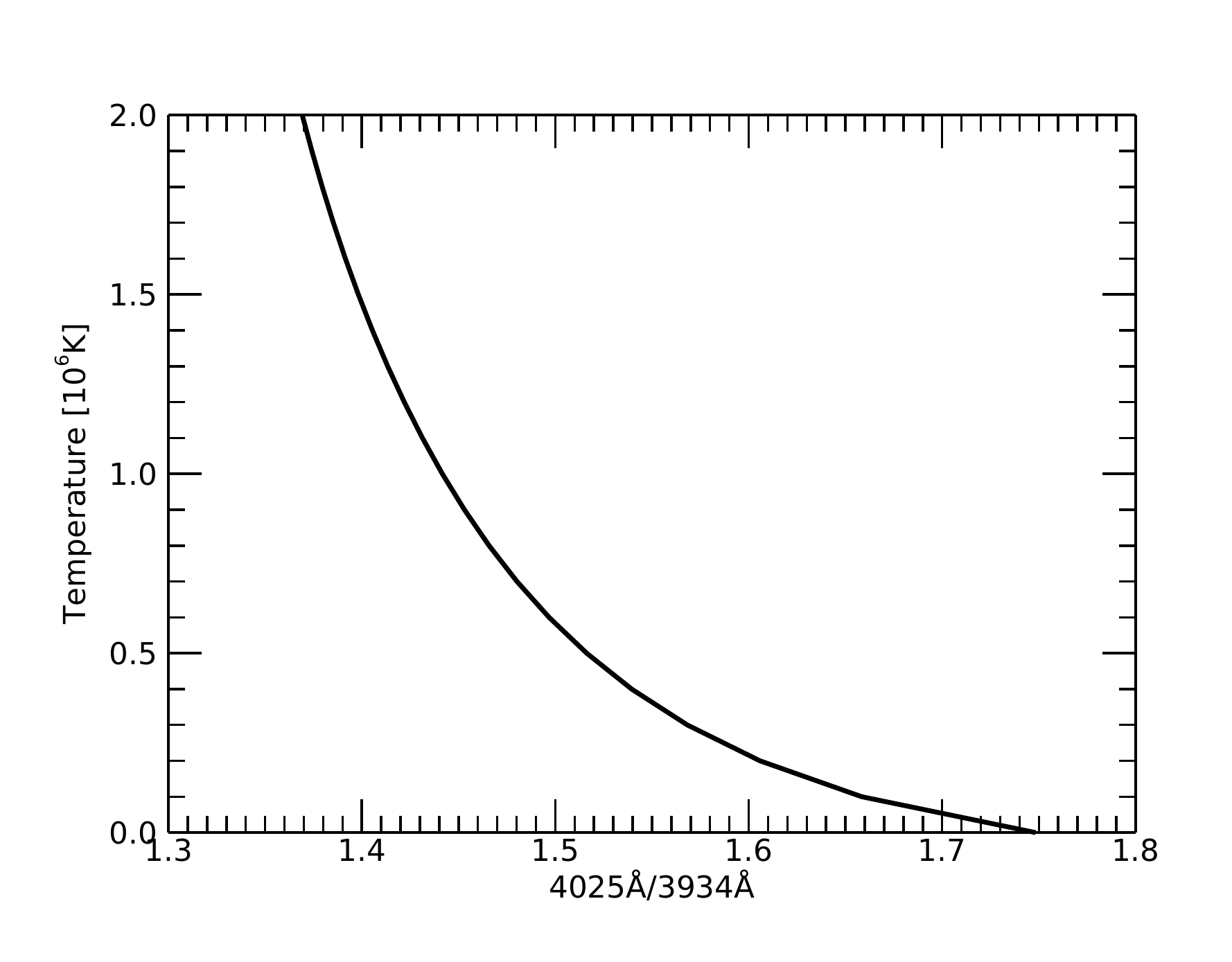}
\caption{The relationship between filter ratio and K-coronal temperature.}
\label{ratiovstemp}
\end{figure}

Electron temperature can be calculated following the anti-correlated relation between the filter ratio and the temperature as shown in Figure \ref{ratiovstemp}.

\begin{figure}[t]
\centering
\includegraphics[width=0.5\textwidth]{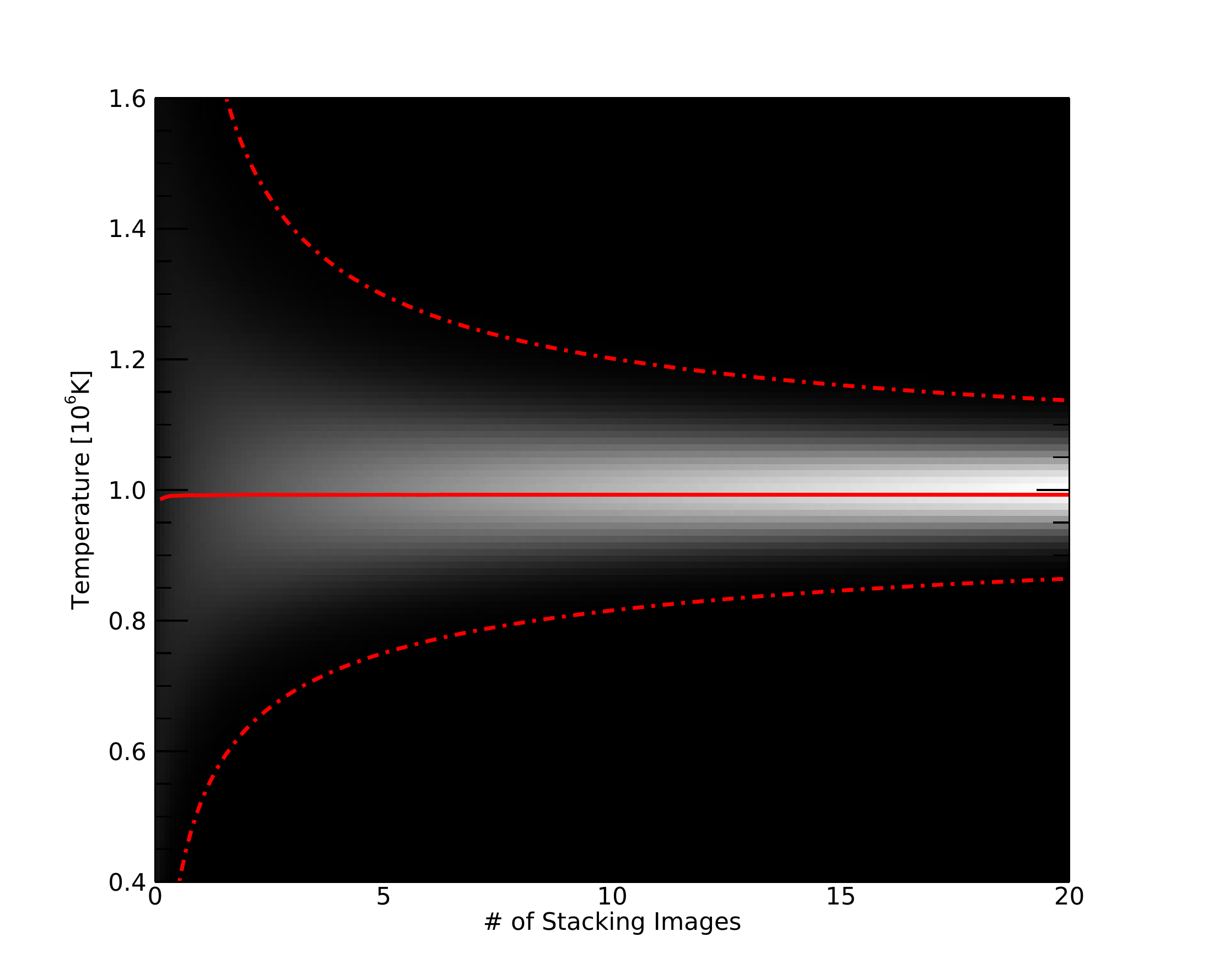}
\caption{Modeling of temperature uncertainty by stacking images after $3\times3$ binning. The simulation performs for 1 MK of coronal temperature, 400 km/s solar wind speed.}
\label{stacking_stddev}
\end{figure}

The deviation of calculated temperature due random noise during the observation is decreased by stacking the images as shown in Figure \ref{stacking_stddev}. Temperature derived from single image with $3\times3$ binning scatters from 0.55\,MK to 1.70\,MK (3-$\sigma$ of temperature). The deviation is decreased by stacking the images. This deviation is mainly caused by the photon noise. Uncertainty of the sky brightness, F-coronal brightness, dark, and readout noise also contribute to it. We expect temperature ranged from 0.77\,MK to 1.26\,MK (about $1\pm0.25$\,MK) if we stack 5 images. Note that, even though the filter ratio follows gaussian distribution, the temperature deviation is asymmetric in y-axis because the function transforming the ratio to temperature is not linear.

\begin{table*}[t]
\centering
\caption{Time of the total solar eclipse and Sun position at the observation site.}
\label{tab:signif_gap_clos}
\begin{tabular}{cccc}
\hline
Observation & Time &  Altitude & Azimuth   \\
            & (UT)         & (degree)           & (degree) \\
\hline
Start of Partial Eclipse  & 16:16:57 & 38.6 & 113.3  \\
Start of Total Eclipse  & 17:35:08 & 50.4 & 134.7  \\
Maximum Eclipse  & 17:36:18 & 50.5  & 135.1  \\
End of Total Eclipse  & 17:37:28 & 50.7 & 135.5 \\
End of Partial Eclipse  & 19:00:42 & 57.8 & 168.6 \\
\hline
\end{tabular}
\end{table*}

\begin{table*}[t]
\centering
\caption{Observation sequence of the DICE}
\label{tab:signif_gap_clos}
\begin{tabular}{cccccc}
\hline
Observation & Filter & Exposure Time & Polarization Angle & Observation Duration & No. of images \\
 & (nm) & (milliseconds) & (degree) & (hh:mm:ss in UT) & \\
\hline
1 & 393.4 & 192 & 0 & 17:35:11 $\sim$ 17:35:16 & 7 \\
 & 393.4 & 192 & 60 & 17:35:17 $\sim$ 17:35:21 & 7 \\
 & 393.4 & 192 & 120 & 17:35:23 $\sim$ 17:35:26 & 6 \\
 & 402.5 & 236 & 0 & 17:35:28 $\sim$ 17:35:31 & 5 \\
 & 402.5 & 236 & 60 & 17:35:33 $\sim$ 17:35:37 & 6 \\
 & 402.5 & 236 & 120 & 17:35:38 $\sim$ 17:35:42 & 6 \\

 & 399.0 & 288 & 0 & 17:35:44 $\sim$ 17:35:47 & 5 \\
 & 399.0 & 288 & 60 & 17:35:49 $\sim$ 17:35:53 & 6 \\
 & 399.0 & 288 & 120 & 17:35:54 $\sim$ 17:35:58 & 6 \\
 & 424.9 & 69 & 0 & 17:36:00 $\sim$ 17:36:04 & 7 \\
 & 424.9 & 69 & 60 & 17:36:05 $\sim$ 17:36:09 & 8 \\
 & 424.9 & 69 & 120 & 17:36:11 $\sim$ 17:36:15 & 8 \\
\hline
2 & 393.4 & 192 & 0 & 17:36:16 $\sim$ 17:36:20 & 6 \\
 & 393.4 & 192 & 60 & 17:36:22 $\sim$ 17:36:25 & 6 \\
 & 393.4 & 192 & 120 & 17:36:27 $\sim$ 17:36:31 & 7 \\
 & 402.5 & 236 & 0 & 17:36:33 $\sim$ 17:36:36 & 6 \\
 & 402.5 & 236 & 60 & 17:36:38 $\sim$ 17:36:42 & 6 \\
 & 402.5 & 236 & 120 & 17:36:43 $\sim$ 17:36:47 & 6 \\
 & 399.0 & 288 & 0 & 17:36:49 $\sim$ 17:36:52 & 5 \\
 & 399.0 & 288 & 60 & 17:36:54 $\sim$ 17:36:57 & 5 \\
 & 399.0 & 288 & 120 & 17:36:59 $\sim$ 17:37:02 & 5 \\
 & 424.9 & 69 & 0 & 17:37:04 $\sim$ 17:37:08 & 7 \\
 & 424.9 & 69 & 60 & 17:37:09 $\sim$ 17:37:12 & 8 \\
 & 424.9 & 69 & 120 & 17:37:14 $\sim$ 17:37:18 & 8 \\
\hline
3 & 393.4 & 192 & 0 & 17:37:20 $\sim$ 17:37:23 & 6 \\
 & 393.4 & 192 & 60 & 17:37:25 $\sim$ 17:37:27 & 4 \\
\hline
\end{tabular}
\end{table*}

%%%%%%%%%%%%%%%%%%%%%%%%%%%%%%%%%%%%%%%%
%%%%%%%%%%%%%%%%%%%%%%%%%%%%%%%%%%%%%%%%

\section{Observation\label{sec:Mission}}

\subsection{2017 TSE experiment }

On August 21, 2017, the total solar eclipse was visible across the entire United States continuously. We observed the eclipse at a site near Jackson hole, Wyoming, where it was anticipated to have good weather. The eclipse began to darken the observation site as the partial eclipse progressed from 16:16:57 UT. The total eclipse began from 17:35:08 UT with 140 seconds duration time. Maximum eclipse occurred at 17:36:18 UT and the solar position was 50.5$^\circ$ of altitude and 135.1$^\circ$ of azimuth. The eclipse ended at 19:00:42 UT. Table 2 shows the eclipse times and positions of the Sun at the observation site during the eclipse.

The observation started about two hours before the total eclipse after aligning the optical system and commissioning the electronics and mechanics of the observation system. Observation modes consist of dark imaging, partial eclipse pinhole imaging, and total eclipse imaging. The dark images were acquired for image preprocessing by shielding the front open aperture of the DICE from light using pinhole cover and dark pouch over the front baffle. One set of dark imaging takes 10 images for 13 levels of exposure time from 0 to 5 seconds, so the total count was 130 images. The pinhole observation was done by a 200$\mu$m pinhole in front of the aperture. The pinhole image is necessary to obtain the brightness of the Sun and the atmospheric transmission of each filter, so a total of 120 images in a set were acquired in 10 repeats of 12 images for four bandwidth filters and three polarization angles. They were taken every 10 minutes during the partial eclipse phases.

Table 3 denotes the total eclipse observation sequence including filter wavelength, exposure time, polarization angle, duration time, and number of images. The exposure times of each sequence for different filters were determined by using the pinhole observation taken one day before the eclipse. Firstly we determined exposure times for pinhole observation of the solar disk. We then estimate the exposure times of the TSE by using assumed disk to corona brightness ratio of 10$^6$ and pinhole to full aperture ratio. Finally we confirmed the exposure times during the partial eclipse observations.

In the filter wheels, there are four filters: 393.4 nm and 402.5 nm for coronal electron temperature sensitive lines, and 399.0 nm and 424.9 nm for speed sensitive lines to determine coronal electron temperature and speed respectively. For the polarizer wheels, there are three polarization angles: 0, 60, and 120 (actually 180, 240, and 300 for 402.5 nm and 424.9 nm filters) degrees to take polarized brightness of the solar corona. Exposure times for each filter (393.4, 402.5, 399.0, and 424.9 nm) are 192, 236, 288, 69 milliseconds, respectively. The spatial pixel size is about 13 arcsec. A total of 162 images were acquired for 136 seconds during TSE from one DICE system. Unfortunately we failed to obtain data from the other CCD camera of the other unit due to a frame transfer error of the camera.

\subsection{Data Analysis} \label{sec:data}

We took flat field images before and after the TSE covering the aperture with white cloth as diffuser, but the obtained images were not consistent. After the observation, we also tried to obtain a flat field image using the dispersed light of the pinhole observation. But we failed to get the flat field because the light passing through the pinhole was not dispersed enough to cover the whole image plane of the CCD. However, the pixel to pixel sensitivity variations and the effect of dust or scratches on the optics and CCD window is expected to be negligible across a few pixels of tracking drift during the TSE. Therefore, the flat field will be cancelled out when we calculate the filter ratio and our result will not be affected. After aligning each filter images, the images were stacked to each polarization to improve the signal to noise ratio.

\begin{figure*}
\includegraphics[width=\textwidth]{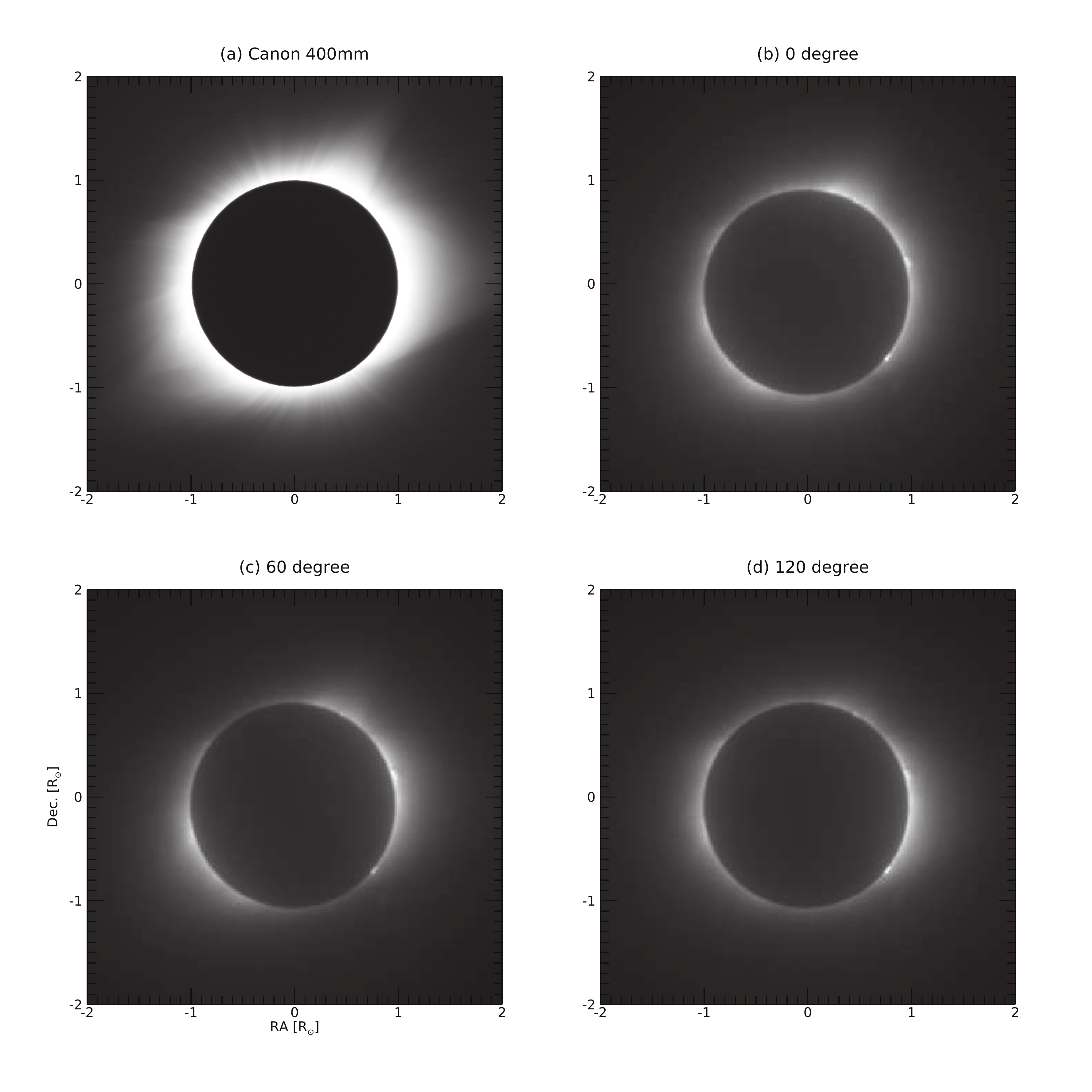}
\caption{WL image and DICE images. The WL image is taken by using Canon 400 mm lens, 2X extender, and camera (upper left) and brightness images taken by using the filter of 402.5.0 nm and the polarizer with three polarization angles.}
\label{wl_dice}
\end{figure*}

Figure 12 shows a white light (WL) image and filtered images during the TSE. The WL image was taken by using the Canon 400 mm lens, 2X extender, and camera with the exposure time of 16 msec. The filter images were taken using 402.5 nm filter for three different polarization angles (0$^\circ$, upper right; 60 $^\circ$, lower left; 120 $^\circ$, lower right). By combining the three polarization angle images in a specific wavelength, we calculate the polarized brightness ($pB_{obs}$), total brightness ($tB_{obs}$), linear polarization degree ($pD_{obs}$), and linear polarization angle ($pA_{obs}$) \citep{billings1966, lu2017}:

\begin{eqnarray}
pB_{obs} &=& \frac{4}{3}[(I_{0^\circ} + I_{60^\circ} + I_{120^\circ})^2\nonumber\\
&& - 3(I_{0^\circ} I_{60^\circ} + I_{60^\circ} I_{120^\circ} + I_{0^\circ} I_{120^\circ})]^{1/2},
\end{eqnarray}

\begin{equation}
tB_{obs} = \frac{2}{3}(I_{0^\circ} + I_{60^\circ} + I_{120^\circ}),
\end{equation}

\begin{equation}
pD_{obs} = \frac{pB_{obs}}{tB_{obs}},
\end{equation}

\begin{equation}
pA_{obs} = \frac{1}{2} \textrm{arccot} \frac{2I_{0^\circ} - I_{60^\circ} - I_{120^\circ}}{\sqrt{3} (I_{60^\circ} - I_{120^\circ})},
\end{equation}

where $I_{0^\circ}$, $I_{60^\circ}$, and $I_{120^\circ}$ are brightness at polarization angles of 0$^\circ$, 60$^\circ$, and 120$^\circ$, respectively.

We adopted pinhole observation with a diameter of 0.2\,mm to correct the difference of the intensity attenuation of the Earth's atmosphere at different wavelengths. The correction factors of the filter ratio are obtained by comparison between the theoretical model intensity and the observed intensity. The theoretical model intensity is calculated by using the instrumental transmissions and the quantum efficiency shown in Figure 8. We used the 2000 ASTM standard extra-terrestrial spectrum reference E-490-00 for the calculation \citep{astm}. The intensity of the pinhole observation of each filter is obtained by integrating the counts in the CCD after dark subtraction. As a result, the atmosphere correction values of about 1.02 and 1.01 for temperature and the speed ratios respectively were obtained.

%%%%%%%%%%%%%%%%%%%%%%%%%%%%%%%%%%%%%%%%
%%%%%%%%%%%%%%%%%%%%%%%%%%%%%%%%%%%%%%%%

\begin{figure*}
\centerline{\includegraphics[height=\textheight]{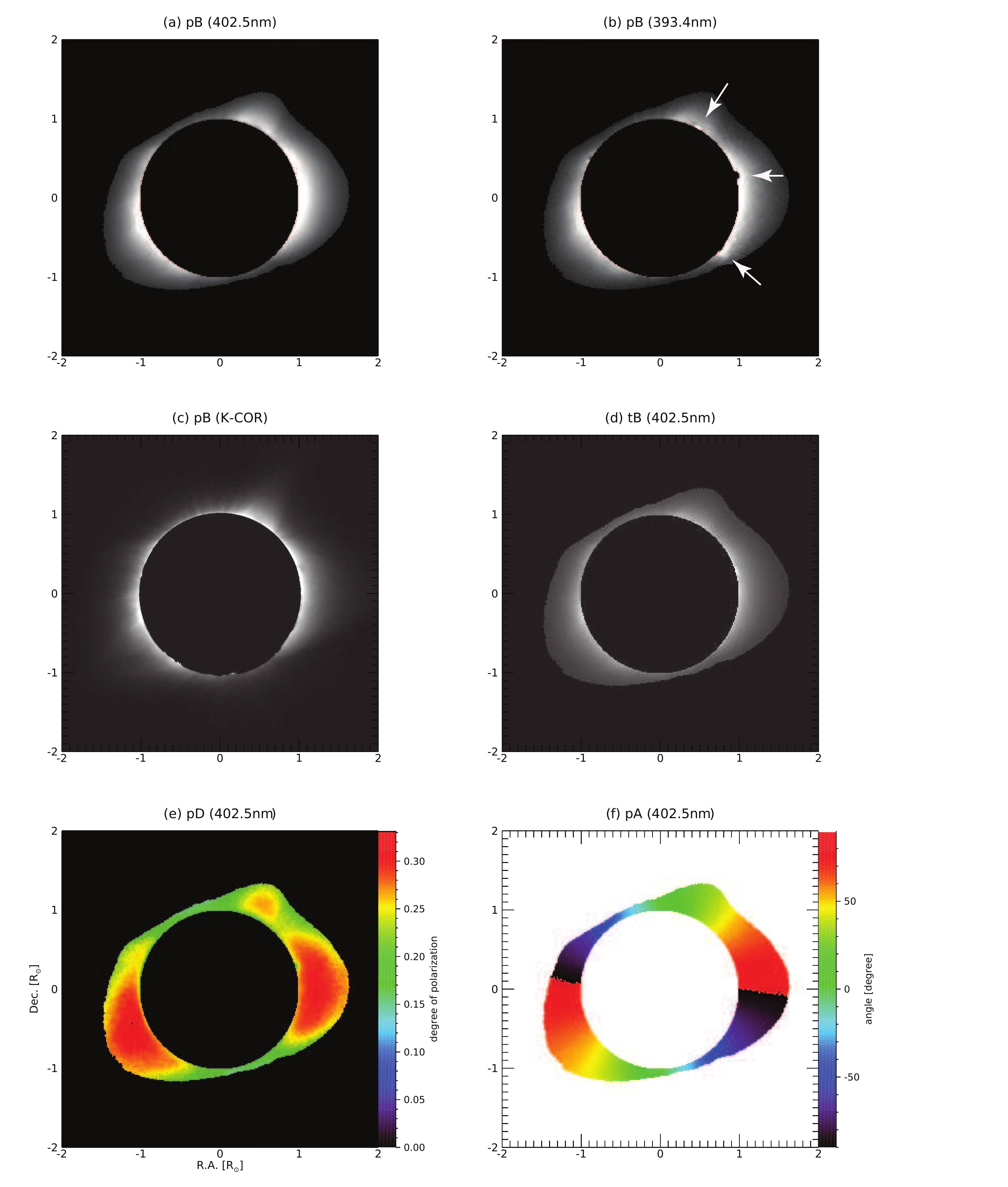}}
\caption{ (a) Polarized brightness of DICE 402.5\,nm filter, (b) 393.4\,nm filter, and (c) K-Cor, (d) total brightness (tB), (e) linear polarization degree (pD), and (f) linear polarization angle (pA) of 402.5\,nm. The DICE images were observed between 17:36:33 UT and 17:36:36 UT, and  K-Cor 735.0 nm were observed at 18:01:52 UT on 21 Aug 2017. The DICE pB and tB are derived from the stacked data of 5 images observed during the total solar eclipse. We mask the region where the DN at 402.5\,nm pB image is lower than 10,000. The white arrows in panel (b) indicate the locations of the stray lights by the prominences. }
\label{img:pbtb}
\end{figure*}

\section{Results}

Figure 13 (a) - (f) show the polarization brightness (pB) at 402.5\,nm, 393.4\,nm, and WL image obtained by MLSO K-coronagraph (K-Cor), the total brightness (tB), degree of linear polarization (pD), and angle of linear polarization (pA) at 402.5\,nm. The pB images of the DICE and the K-Cor show that the coronal structures such as streamers located at the east, west, and north-west, and polar plumes located at the north pole look almost similar to each other. It looks likely that the DICE observed the coronal structures and acquired their polarization information successfully. From this images, we find the following characteristics: (1) Coronal structures such as streamers at the east, west, and north-west are well observed both in pB and tB images. (2) Fine threaded structures appear more clearly in the pB image than in the tB image. (3) Polar plumes at the north pole are identified in the pB image.

We note that the pB at 393.4\,nm shows radial leaking pattern from prominences located at the west limb as denoted by arrows in Figure 13 (b). The raw images at 393.4\,nm was saturated and contaminated by strong scattering by the light from the prominences. The band pass of 393.4\,nm filter includes the strong chromospheric lines such as Calcium H \& K centered at $\lambda$ = 396.8 nm and 393.3 nm.

\begin{figure*}
\includegraphics[width=0.5\textwidth]{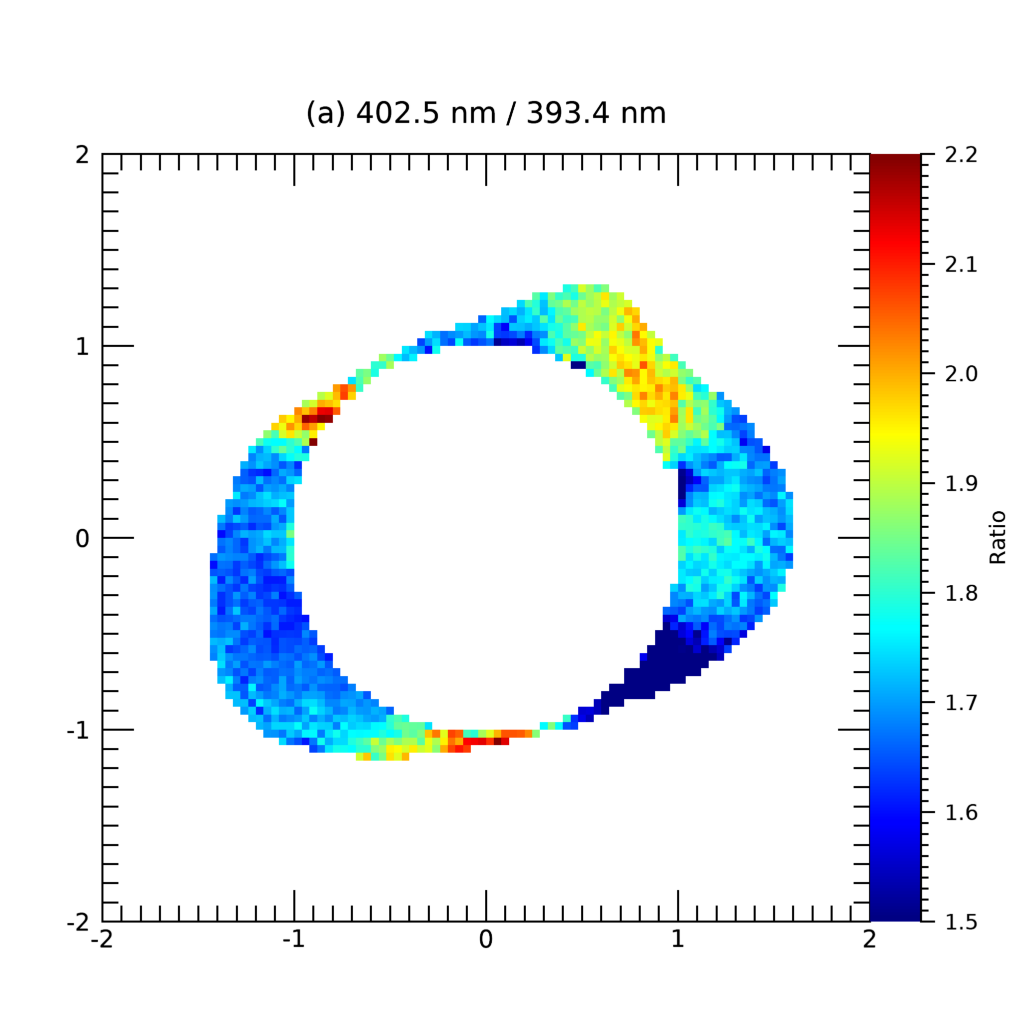}
\includegraphics[width=0.5\textwidth]{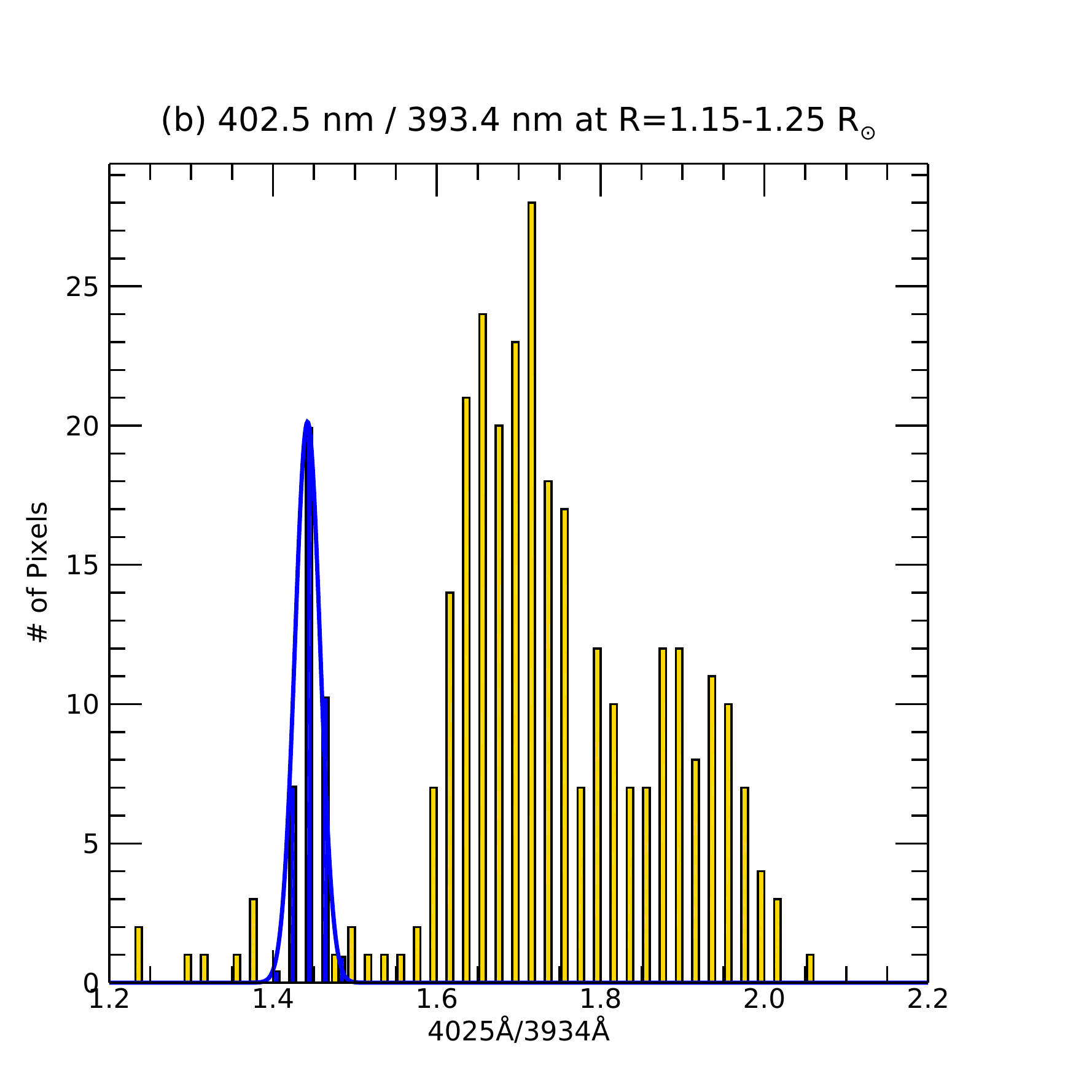}
\caption{The ratio of 402.5\,nm and 393.4\,nm filter images for temperature measurement and the distribution of the ratio at $R=1.15-1.25R_\odot$. Yellow histogram represents the distribution of observed value and the blue histogram denotes the simulated distribution at the same region. \label{fig:ratio_map}}
\end{figure*}

To deduce coronal information, we only consider temperature sensitive wavelengths. We don't consider the speed sensitive lines because the solar wind in the DICE field-of-view has a low speed ($\sim$ 10 km s$^{-1}$) that is much less than the uncertainties ($\sim$ 100 km s$^{-1}$) of the speed estimation from TSE observation.
It is well accepted that the solar wind acceleration region is in the heliocentric distance range 2 $\sim$ 8 R$_\odot$ \citep {antiochos2011}.
Figure \ref{fig:ratio_map} shows a filter ratio map of 402.5\,nm and 393.4\,nm filters and a histogram of the observed (yellow bars) and the simulated filter ratio (blue bars) at $R=1.15-1.25R_\odot$.
As the atmospheric transmission may be different in each filter band, the observed filter ratio was corrected using the pinhole observation data as described in previous section.
From the data, the ratio is distributed in the range of 1.5 - 2.0. The ratio map reflects some patterns of the observed filter images, such as the boundary of the streamer on the west limb, threaded polar plume on the north limb. One can also notice the same radial leaking pattern from prominences located at the west limb seen in Figure 13 (b). Most of the pixels on the west limb are likely affected by this stray light because three bright prominences are located there. We note that the stray light seems to depend on the wavelength as comparing the four filter images.

The distribution of the simulated value in the Figure \ref{fig:ratio_map} (b) represents the filter ratio for temperature measurement after combining 5 images per each polarization. The observed value is centered at around 1.73 and scattered from 1.2 to 2.25, while the simulated value is centered at 1.44.
The discrepancy can not be explained with the systematic tolerance ($3\sigma\sim$0.15) investigated in the section \ref{sec:tol}, which implies that other unknown error sources are involved.

%\textbf{In the simulation we assume 1 MK as a mean temperature of the corona.}

 \begin{figure*}
\includegraphics[width=0.5\textwidth]{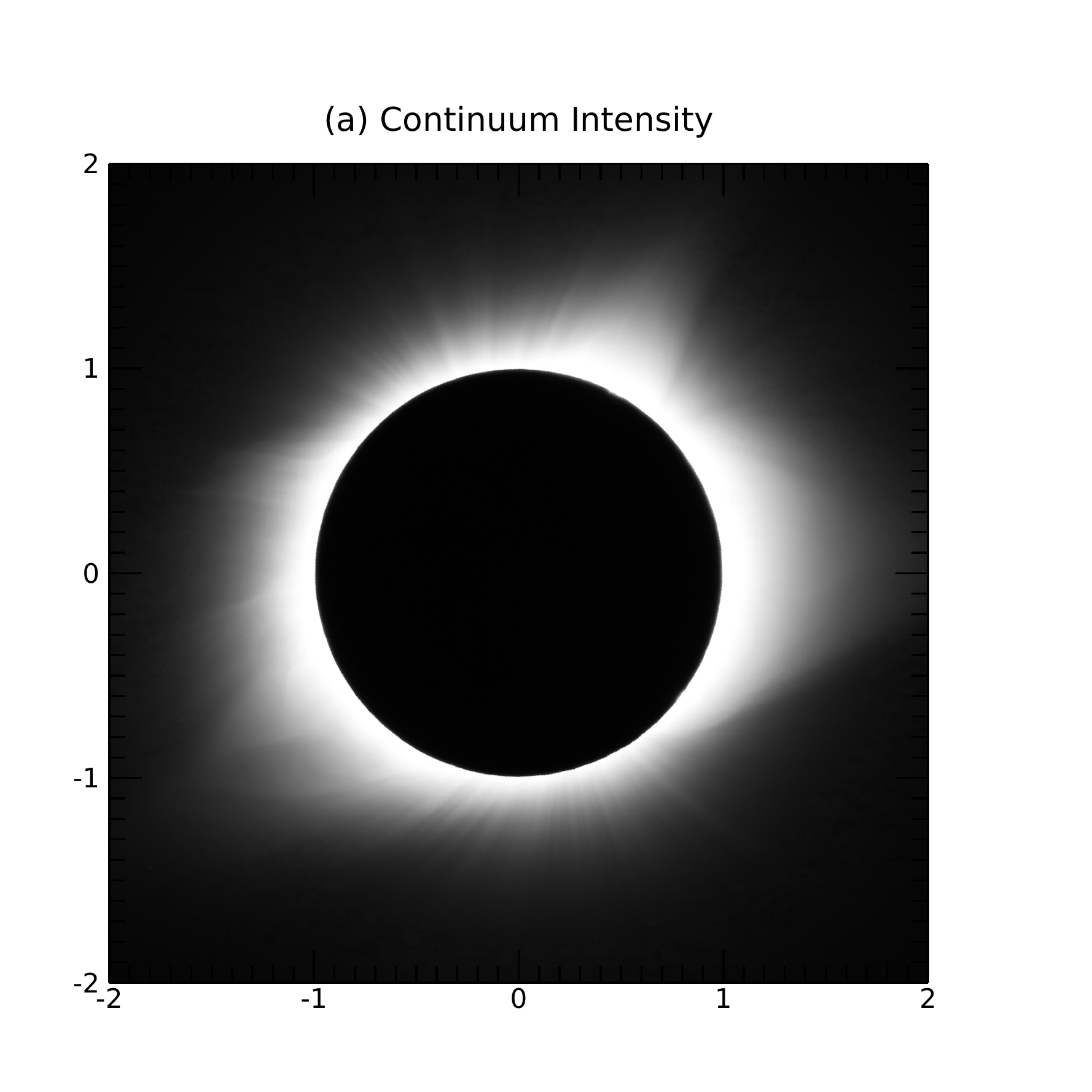}
\includegraphics[width=0.5\textwidth]{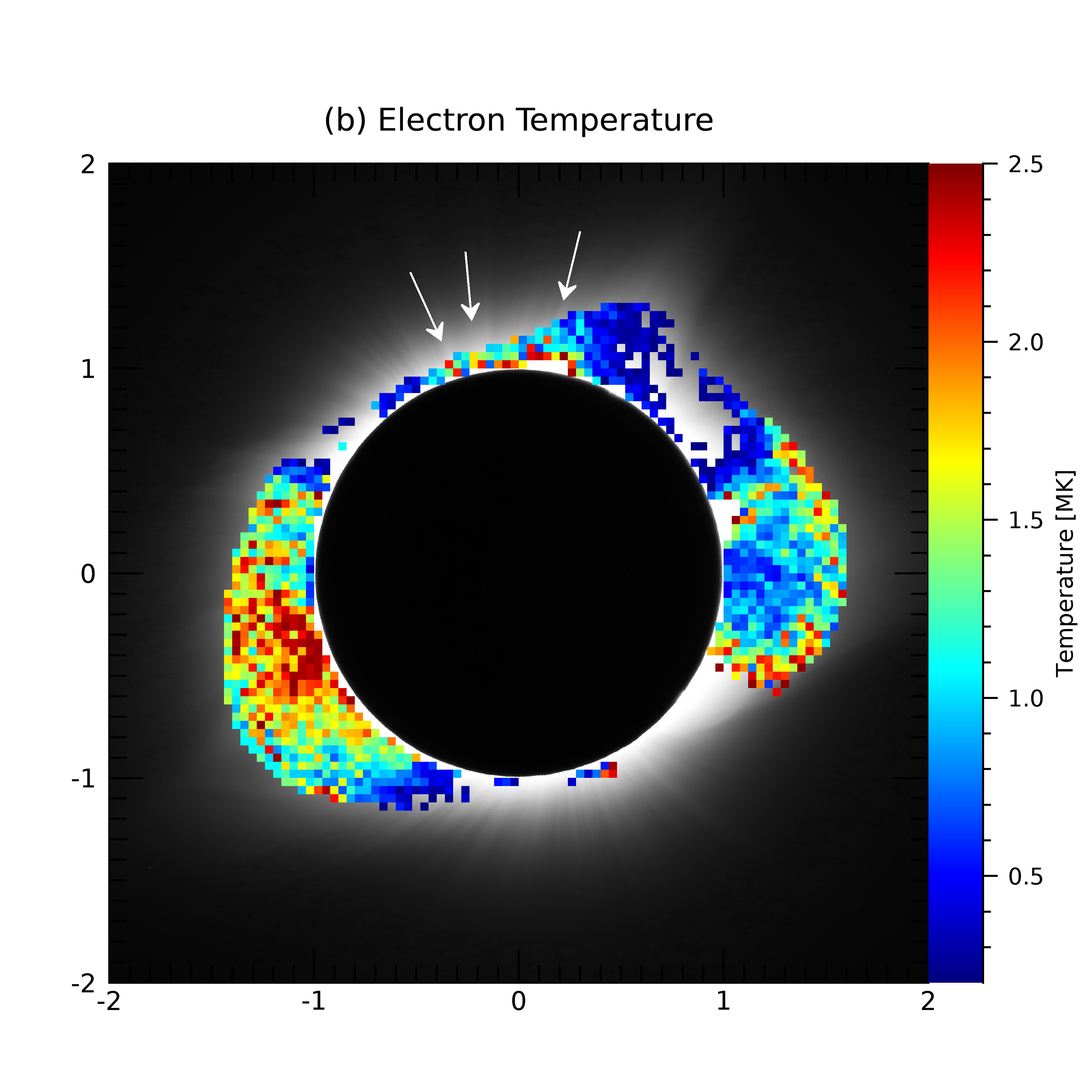}
\caption{Coronal continuum intensity image (left) and distribution of electron temperature (right) over the intensity image. \label{fig:temp_map}}
\end{figure*}

For this reason, we adopted a correction factor, which is not used to get an exact temperature value of corona but a tendency of temperature structure. By comparison between the theoretical model intensity ratio and observed intensity ratio, we obtained the correction factor of 0.83. We applied the value of 0.83 to the observed filter ratio, and we then converted the filter ratio to temperature value by using the relationship between filter ratio and K coronal temperature, as shown in Figure 10. Finally, we derived the electron temperature ranging from $0.2 - 2.5$\, MK as shown in Figure 15. In the Figure, the temperature map is displayed over a continuum intensity image observed during the total solar eclipse.

We cannot define exact values of the temperature from the derived temperature map, because west limb is contaminated by the stray light and we cannot fully understand the correction factor. However, the tendency of the temperature on the east limb of the corona is acceptable, since the temperature of the streamer on the east limb is higher than the surroundings, and its boundaries between the streamer and the cold polar coronal hole can be identified clearly.

The threaded polar plumes are likely to have high temperature compared to surroundings as marked with the white arrows in Figure 15.

\section{Discussion \label{sec:discussion}}

To verify fundamental technology of the NGC, we developed the DICE that can measure the coronal intensity through filter and polarizer wheels. We then  conducted the TSE observation on 2017 Aug 21 at Jackson Hole, Wyoming. The DICE instrument successfully obtained the polarization information such as polarization brightness, angle, and degree of solar corona after stacking five images in each polarization. Using the ratio of polarization brightness of 402.5\,nm and 393.4\,nm filter, we extract the temperature distribution map of the corona as a main result.

The estimated temperature seems consistent with known trends in the east limb. It is higher at the streamer and lower at its boundaries between the streamer and cool polar corona hole. We can also see that several structures have a consistent shape along the thread of the streamer.

The west limb and south region of our observation did not show proper temperature range, probably due to the contamination by light scattering from prominences. The result of the Zemax optical simulations showed that scattering converged to zero, so we set the scattering to zero in the simulation of section 3.4. However, the real value is probably large and it may be a function of wavelength. If the scattering changes with the filter images, the ratio value would be affected. Moreover, lens flare can be generated by a strong light source such as prominences. The lens flares were seen at the three prominences located at the west limb. In particular, the wavelength band of the filter centered at 3934\,\AA~contains Ca {\sc ii} H / K lines. The brightness of the prominence is comparable to the disk brightness, so that the lens flares generated by the prominence can be much brighter than the corona where brightness is 10$^6$ times fainter than the disk. The scattering by the prominences appears to have radically changed not only the filter ratio around the prominence, but also the ratio at the overall west limb.

The estimated temperature is obtained by the $3\times3$ binning and stacking 5 images. The Monte Carlo simulation shows a large temperature deviation even with $3\times3$ binning without image stacking. We expect the temperature deviation of about 0.3\,MK when we stack the five set of the images obtained in the observing run. However, the variation of the measured ratios is much larger than the expected variation. There is a difference in the ratios between the simple symmetric density model that we used in the temperature calculation and asymmetric model considering the direction of streamers (front side or backside) as shown in \citet{reginald2004} and \citet{ichimoto1996}. However, the difference by the models is not big enough to explain the significant difference between the simulation and the observation. It looks likely that the difference may be due to other sources not properly assessed in systematic tolerance analysis, such as stray light by the prominence, diffraction pattern in pinhole images causing error in atmospheric attenuation, lack of proper flat field image, or other still unknown factors we missed. Further studies should be followed to resolve the issue.

Through the experiment, we obtained several lessons for the next generation coronagraph. First, we could get significant benefits to the development of the control software by using core Flight System (cFS), which is NASA open source for embedded systems. The cFS is already technically proven and easy to apply to a portable telescope system. Second, we could obtain the temperature map of the low corona in the eastern limb, although the temperature measurement of the DICE has an offset to the theoretical simulation. However, we could not achieve the information on the western limb due to unexpected strong emissions by the prominences. We also could not get enough data to increase the signal to noise ratio because one of the cameras did not work during the eclipse. From this experience, we learned the importance of lab experiments for the whole system. Basically, we needed to make sure that the optics and filters meet component and system-level requirements. For component-level testing, the manufacturer was asked to provide test results, and the lab checked the system-level requirements to ensure that the system reached the specified system performance. In the lab testing, the resolution requirement was tested by measuring a spot size with an on-axis white collimated light source. The camera issue of DICEs may come from the direct power usage of a car battery. A power distribution system is especially vital to a camera to guarantee reliable operation during time-critical observations. Our TSE observation was extended to a new technology demonstration experiment called BITSE (Balloon-borne Investigation of Temperature and Speed of Electrons in the Corona). It will be extended to the CODEX (COronal Diagnostic EXperiment). The follow-up NASA-KASI joint mission will enable us to understand the coronal heating and solar wind acceleration by obtaining the electron density, temperature, and velocity simultaneously.

%%% ACKNOWLEDGMENTS (IF ANY) %%%%%%%%%%%%%%%%%%%%%%%%%%%%%%%%%%%%%%%%

\acknowledgments
%We are very thankful to Prof. Valery Nakariakov and Prof. Bo Li for their constructive comments and helpful suggestions on the future science.
This work was supported the Korea Astronomy and Space Science Institute under the R\&D program (2020-1-850-02) supervised by the Ministry of Science, ICT and Future Planning,

\end{document}